\documentclass[amsmath,amssymb,floatfix,reprint,twocolumn,superscriptaddress]{revtex4-1}

\usepackage{graphicx}
\usepackage{epsfig}
\usepackage{hyperref}

\usepackage{xcolor}

\begin{document}

\title{Bulk and element specific magnetism of the medium and high entropy Cantor-Wu alloys}

\author{D.~Billington}
\altaffiliation[Current address: ]{School of Physics and Astronomy, Cardiff University, Queen's Building, The Parade, Cardiff, CF24 3AA, United Kingdom}
\email{billingtond1@cardiff.ac.uk}
\affiliation{H.H.~Wills Physics Laboratory, University of Bristol, Tyndall Avenue, Bristol, BS8 1TL, United Kingdom}
\affiliation{School of Physics and Astronomy, Cardiff University, Queen's Building, The Parade, Cardiff, CF24 3AA, United Kingdom}
\affiliation{Japan Synchrotron Radiation Research Institute, SPring-8, Sayo 679-5198, Japan}
\author{A.D.N.~James}
\affiliation{H.H.~Wills Physics Laboratory, University of Bristol, Tyndall Avenue, Bristol, BS8 1TL, United Kingdom}
\author{E.I.~Harris-Lee}
\affiliation{H.H.~Wills Physics Laboratory, University of Bristol, Tyndall Avenue, Bristol, BS8 1TL, United Kingdom}
\author{D.A.~Lagos}
\affiliation{H.H.~Wills Physics Laboratory, University of Bristol, Tyndall Avenue, Bristol, BS8 1TL, United Kingdom}
\author{D.~O'Neill}
\affiliation{Department of Physics, University of Warwick, Coventry, CV4 7AL, United Kingdom}
\author{N.~Tsuda}
\affiliation{Japan Synchrotron Radiation Research Institute, SPring-8, Sayo 679-5198, Japan}
\author{K.~Toyoki}
\affiliation{Japan Synchrotron Radiation Research Institute, SPring-8, Sayo 679-5198, Japan}
\author{Y.~Kotani}
\affiliation{Japan Synchrotron Radiation Research Institute, SPring-8, Sayo 679-5198, Japan}
\author{T.~Nakamura}
\affiliation{Japan Synchrotron Radiation Research Institute, SPring-8, Sayo 679-5198, Japan}
\author{H.~Bei}
\affiliation{Materials Science and Technology Division, Oak Ridge National Laboratory, Oak Ridge, TN 37831, USA}
\author{S.~Mu}
\affiliation{Materials Science and Technology Division, Oak Ridge National Laboratory, Oak Ridge, TN 37831, USA}
\author{G.D.~Samolyuk}
\affiliation{Materials Science and Technology Division, Oak Ridge National Laboratory, Oak Ridge, TN 37831, USA}
\author{G.M.~Stocks}
\affiliation{Materials Science and Technology Division, Oak Ridge National Laboratory, Oak Ridge, TN 37831, USA}
\author{J.A.~Duffy}
\affiliation{Department of Physics, University of Warwick, Coventry, CV4 7AL, United Kingdom}
\author{J.W.~Taylor}
\affiliation{DMSC - European Spallation Source,  Universitetsparken 1, Copenhagen 2100, Denmark}
\author{S.R.~Giblin}
\affiliation{School of Physics and Astronomy, Cardiff University, Queen's Building, The Parade, Cardiff, CF24 3AA, United Kingdom}
\author{S.B.~Dugdale}
\affiliation{H.H.~Wills Physics Laboratory, University of Bristol, Tyndall Avenue, Bristol, BS8 1TL, United Kingdom}

\date{\today}

\begin{abstract}
Magnetic Compton scattering, x-ray magnetic circular dichroism spectroscopy and bulk magnetometry measurements are
performed on a set of medium (NiFeCo and NiFeCoCr) and high (NiFeCoCrPd and NiFeCoCrMn) entropy Cantor-Wu alloys.
The bulk spin momentum densities determined by magnetic Compton scattering are remarkably isotropic, and this is
a consequence of the smearing of the electronic structure by disorder scattering of the electron quasiparticles.
Non-zero x-ray magnetic circular dichroism signals are observed for every element in every alloy indicating differences in
the populations of the majority and minority spin states implying finite magnetic moments.
When Cr is included in the solid solution, the Cr spin moment is unambiguously antiparallel to the total magnetic moment,
while a vanishingly small magnetic moment is observed for Mn, despite calculations indicating a large moment.
Some significant discrepancies are observed between the experimental bulk and surface magnetic moments.
Despite the lack of quantitative agreement, the element specific surface magnetic moments seem to be qualitatively reasonable.
\end{abstract}

\maketitle

\section{Introduction}

It has been known for centuries that combining elements together in an alloy can result in the material having 
superior properties. Traditionally, however, there would be one principal component ({\it e.g.} Fe) to which smaller amounts of
other elements ({\it e.g.} C) are added. Indeed, the Bronze Age is characterized by the technological advance enabled by the discovery
that adding small amounts of other elements (such as Sn) to Cu produced a harder metal.
Building on early work published mainly in undergraduate theses (see, for example \cite{cantor:14}), 
several publications \cite{cantor:04,chen:04,hsu:04,yeh:04,yeh:04a} established in 2004 the existence of a new type of alloy
that is formed not by adding small amounts of other elements to one principal component, but by combining several elements in approximately
equiatomic proportions.
These alloys are often referred to as `high entropy alloys' (HEAs)
\cite{tsai:13,tsai:14,zhang:14,kozak:15,pickering:16,miracle:17}
which was a term first coined by Yeh {\it et al.} who attributed the
high configurational entropy as the mechanism stabilising the solid
solution phase \cite{yeh:04a}. The terms `multicomponent alloys' and `multiprincipal element alloys' are also commonly used.

HEAs have complete substitutional disorder meaning that all of the component elements in the material randomly occupy
the crystallographic sites and, as such, these alloys do not have long range compositional order (although there is growing evidence
that short range order can exist in HEAs \cite{lucas:12,zhang:14,tamm:15,pickering:16,miracle:17,zhang:17,dong:19,zhang:20}).
This degree of disorder introduces unusual and unexpected behavior including enhanced mechanical
properties such as hardness and resistance to wear, tensile strength, ductility and fracture
resistance  \cite{pickering:16,kozak:15,tsai:13,zhang:14,miracle:17,gludovatz:14}. 
These enhanced mechanical properties have seen high entropy alloys become candidate materials for
potential next generation engineering applications including use in state-of-the-art racing cars,
spacecraft, submarines, jet aircraft and nuclear reactors \cite{senkov:16,csanadi:19,barron:20}.

Shortly after Yeh named these alloys, Cantor {\it et al.} developed the prototypical, equiatomic
high entropy alloy NiFeCoCrMn \cite{cantor:04}, known as the `Cantor alloy', 
which has been the subject of considerable work in the field.
More recently, Wu {\it et al.} \cite{wu:14} showed that alloying the individual elements of Cantor's NiFeCoCrMn alloy
with each other and with Pd produced a series of two, three and four component equiatomic fcc
solid solutions collectively referred to as `Cantor-Wu' alloys \cite{cantor:04,wu:14} that include
NiPd, NiCo, NiFe, NiFeCo, NiCoCr, NiCoMn, NiCoCrMn, NiFeCoMn, NiFeCoCr, NiFeCoCrMn and NiFeCoCrPd.
Currently, there is not a universal classification system which exactly qualifies an alloy to be a HEA,
but alloys with five or more elemental components with this high substitutional disorder are generally considered to be HEAs
\cite{lucas:12,tsai:13,tsai:14,zhang:14,pickering:16,kozak:15,niu:15,miracle:17}
while alloys containing few components are given the appellation `medium entropy alloys'. 
Although much of the interest in HEAs stems from their potential for use in industrial and technological applications,
from a fundamental physics perspective the Cantor-Wu alloys display a rich variety of electronic and magnetic behavior.

\begin{table}[t!]
\caption{Previously reported experimental Curie temperatures, $T_{\rm C}$, spin freezing temperatures,
$T_{\rm f}$, Kondo temperatures, $T_{\rm K}$, and saturated magnetic moments, $m^{\rm sat}$,
for selected Cantor-Wu alloys taken from Refs.~\cite{jin:16,jin:17,sales:16,kao:11,schneeweiss:17,lucas:11}.}
\centering 
\begin{tabular}{c c c c c} 
\hline\hline 
Alloy & $T_{\rm C}$ & $T_{\rm f}$ & $T_{\rm K}$ & $m^{\rm sat}$ \\
      & (K) & (K) & (K) & ($\mu_{\rm B}$~atom$^{-1}$) \\
\hline 
NiFeCo     \cite{jin:16,jin:17}         & $995$  & -    & -    & $1.7$  \\
NiFeCoCrPd \cite{jin:16,lucas:11}       & $440$  & -    & -    & $0.52$ \\
NiFeCoCr   \cite{jin:16,kao:11}         & $120$  & $35$ & -    & $0.24$ \\
NiCoCr     \cite{jin:16,sales:16}       & $<2$   & -    & -    & $0$    \\
NiFeCoCrMn \cite{jin:16,schneeweiss:17} & $38$   & $93$ & $40$ & $<0.01$       \\
\hline \hline
\end{tabular}
\label{tab}
\end{table}

The Cantor-Wu alloys represent a mixture of $3d$ transition metal ions and it is well known that magnetism (specifically, the $d$-band filling)
in the $3d$ transition metals is responsible for both their particular ground-state crystal structures and their mechanical properties \cite{pettifor:88,guo:11}.
So far, experimental information regarding the magnetism comes from bulk magnetometry measurements
\cite{zhang:13,jin:16,sales:16,kao:11,schneeweiss:17,lucas:11} which only provide the
total (the sum of spin and orbital) magnetic moments.
The only available element specific information about the magnetism comes from {\it ab initio} calculations
\cite{zhang:13,tian:13,jin:16,schneeweiss:17,mu:19,mu:19a,lagosphd}.
The combination of bulk magnetometry measurements and calculations has revealed that most of
the Cantor-Wu alloys studied here are either ferromagnetic or ferrimagnetic.
NiFeCoCr and NiFeCoCrMn have also been reported to exhibit spin glass behavior \cite{kao:11,jin:16,schneeweiss:17},
and Kondo-like behavior has been observed in NiFeCoCrMn as evidenced by an upturn in the
resistivity at low temperatures.
Table~\ref{tab} lists the previously reported experimental values of the Curie
temperatures, $T_{\rm C}$, and (where relevant) spin freezing temperatures,
$T_{\rm f}$, and the Kondo temperatures, $T_{\rm K}$, together with their
saturated magnetic moments, $m^{\rm sat}$ \cite{jin:16,sales:16,kao:11,schneeweiss:17,lucas:11}.
Except for NiFeCoCrMn, the Curie temperatures and saturated magnetic moments decrease with increasing Cr concentration
and it has been argued that this implies a Cr moment aligned antiparallel to the average total moment,
as predicted by calculations \cite{jin:16,schneeweiss:17,mu:19,mu:19a,lagosphd}.

The presence of strong compositional disorder and magnetism in these alloys results in non-trivial electronic transport properties.
Residual resistivity measurements show that the Cantor-Wu alloys are split into two subgroups with low
($<10$~$\mu\Omega\cdot$cm) and high ($>75$~$\mu\Omega\cdot$cm) residual resistivities \cite{zhang:15,jin:16}.
Interestingly, the members of these two groups are not determined by the number of component elements,
but instead are determined by the type of elements present.
Korringa–Kohn–Rostoker (KKR) calculations employing the coherent potential approximation (CPA),
which can effectively treat the compositional disorder, indicate substantial smearing of the electronic
structure due to scattering of the electron quasiparticles \cite{zhang:15,samolyuk:18,mu:18,mu:19,mu:19a}.
For alloys containing only Fe, Co and/or Ni, the majority spin channel experiences negligible disorder scattering
thereby providing a short circuit, while for Cr/Mn containing alloys both spin channels experience strong
disorder scattering due to an electron filling effect \cite{mu:19}.
Very recently, it was found experimentally that NiFeCoCr has a quasiparticle coherence length that is very close
to the nearest neighbor interatomic distance \cite{robarts:20}, {\it i.e.} approaching the Mott-Ioffe-Regel limit
where the standard picture of ballistically propagating quasiparticles becomes invalid \cite{hussey:04}.
In fact, NiCoCr, NiFeCoCrMn and NiFeCoCrPd all have residual resistivities which are higher than NiFeCoCr,
and all exhibit non-Fermi liquid behavior \cite{jin:16} and should probably be classed as trivial non-Fermi liquids \cite{buterakos:19}.
Quantum critical behavior has been reported in NiCoCr$_{x}$ ($x\approx1$) whose
magnetic moment vanishes due to strong magnetic fluctuations \cite{sales:16}.
Despite all of these observations, there is currently no element specific experimental information about the
magnetic moments of each alloy.

In this study, we report magnetic field dependent synchrotron x-ray experiments with circularly polarized photons
and bulk magnetometry for a set of medium (NiFeCo and NiFeCoCr) and high (NiFeCoCrPd and NiFeCoCrMn) entropy
Cantor-Wu alloys \cite{cantor:04,wu:14}.
Magnetic Compton scattering \cite{cooper:97,mccarthy:97} is used to determine the magnetic Compton profiles (MCPs) along
high symmetry crystallographic directions.
The MCPs are one dimensional projections of the underlying three dimensional bulk spin momentum density
which is intimately related to the many body ground state electronic wavefunction.
Magnetic Compton scattering can determine the bulk spin moment and can be used to determine the bulk orbital moment by subtracting the bulk spin moment from the total bulk moment determined,
for example, by bulk magnetometry measurements.
X-ray magnetic circular dichroism (XMCD) \cite{funk:05} spectroscopy at the $L_{2,3}$-edges of the $3d$-elements
($M_{2,3}$-edges of Pd) is exploited to obtain element specific orbital and spin magnetic moments via the orbital
and spin sum rules \cite{thole:92,carra:93}, and to track their variation with applied magnetic field.

Before proceeding, we would like to emphasize that we do not expect the XMCD orbital and spin sum rules to provide
quantitatively accurate values for the spin and orbital moments of the Cantor-Wu alloys due to the inherent surface
sensitivity of total electron yield detection, the maximal compositional disorder of the measured alloys, and
uncertainties in quantities that enter the sum rule equations, such as the element specific $d$-electron occupancy
of each alloy and the expectation value of the magnetic dipole operator at the surface of the samples.
Despite this, the XMCD spectra will unambiguously reveal whether there is a finite moment and whether it is aligned parallel
or antiparallel to the applied field.
Furthermore, we expect the relative sizes of the sum rule moments to be qualitatively correct.

\section{Methods}

\subsection{Crystal growth}

Details of the single crystal growth can be found in Refs.\cite{bei:05,wu:14}.
Ingots of each alloy were produced by arc melting the constituent elements in a water cooled copper hearth,
under an Ar atmosphere.
The arc melted buttons were flipped and remelted five times in order to improve
the compositional homogeneity, before being drop cast into square cross section copper moulds.
These polycrystalline ingots were then loaded into an optical floating zone furnace to produce a
single crystal which was subsequently cut into a disc using electrodischarge machining before being
electrolytically polished to remove any damage caused by the cutting.

\subsection{Sample preparation}

For the Compton scattering measurements,
the samples were aligned along high-symmetry crystallographic directions using x-ray Laue diffraction.
The fcc crystal symmetry was evident in the recorded diffraction patterns.

In order to remove the contaminated surface oxide layer for the x-ray absorption measurements,
all of the samples were chemically etched for $3$~min in a solution of distilled H$_{2}$O, $37$~wt.\% HCl,
and $70$~wt.\% HNO$_{3}$ with a $\mathrm{H_{2}O}:\mathrm{HCl}:\mathrm{HNO_{3}}$ volume ratio of $1:2:1$.
After etching, the samples were introduced to the ultrahigh vacuum (UHV) chamber of the soft x-ray absorption
spectrometer apparatus.
The samples were then sputtered {\it in situ} with an Ar ion plasma.
For the Ar ion sputtering, the acceleration voltage was $2$~kV, the emission current was $10$~mA,
the incident angle was $45^{\circ}$, the Ar pressure was $6.7\times10^{-3}$~Pa,
and the duration was $2$~h.
There was no sample alignment for the x-ray absorption measurements, partly due to the sample preparation procedure, but also because cubic systems necessarily have low magnetocrystalline anisotropy.

\subsection{Bulk magnetometry measurements}

The magnetic field dependence of the total bulk magnetic moments, $m^{\rm tot}_{z}(H_{\rm ext})$,
of NiFeCo, NiFeCoCr, NiFeCoCrPd and NiFeCoCrMn were measured at $T=10$~K using a
superconducting quantum interference device (SQUID).
All of the samples were cooled in zero field.

\subsection{Magnetic Compton scattering measurements}

The electron momentum density from Bloch electrons can be written as,
\begin{equation}
\begin{split}
\rho({\bf p}) & =\sum_{j,{\bf k}}n_{{\bf k},j}\bigg|\int\psi_{{\bf k},j}({\bf r})\exp(-\text{i}{\bf p}\cdot{\bf r})~\text{d}^{3}{\bf r}\bigg|^{2}\\
              & =\sum_{j,{\bf k},{\bf G}}n_{{\bf k},j}\big|a_{{\bf G},j}({\bf k})\big|^{2}\delta({\bf p}-{\bf k}-{\bf G}),
\end{split}
\end{equation}
where $\psi_{{\bf k},j}({\bf r})$ is the real-space wavefunction of an electron in band $j$ with wavevector ${\bf k}$,
$n_{{\bf k},j}$ is its occupancy which takes values between $0$ (unoccupied) and $1$ (fully occupied),
and the $\delta$-function expresses the contribution from higher momentum (Umklapp) components whose
intensities are given by the Fourier coefficients of the real-space electron wavefunctions, $a_{{\bf G},j}({\bf k})$.
This means that an occupied electron state will contribute to the electron momentum density not only at ${\bf p}={\bf k}$
but also at ${\bf p}={\bf k}+{\bf G}$, where ${\bf G}$ is {\it any} reciprocal lattice vector.

The Compton profile, $J(p_{z})$, is defined as the one dimensional (twice integrated) projection of the
electron momentum density, $\rho({\bf p})$, along the scattering vector which is parallel
(orthogonal) to $p_{z}$ ($p_{x,y}$),
\begin{equation}
J(p_{z})=\iint\rho({\bf p})~\text{d}p_{x}\text{d}p_{y},
\end{equation}
which is normalized to the number of electrons, $N$,
\begin{equation}
N=\int J(p_{z})~\text{d}p_{z}.
\end{equation}
It is also possible to write the total electron momentum density as the sum of contributions from
the momentum densities of electrons with spins aligned parallel, $\rho_{\uparrow}({\bf p})$,
or antiparallel, $\rho_{\downarrow}({\bf p})$, to a chosen spin quantization axis (the scattering vector),
\begin{equation}
\rho({\bf p})=\rho_{\uparrow}({\bf p})+\rho_{\downarrow}({\bf p}),
\end{equation}
and we can then define the electron spin momentum density as,
\begin{equation}
\rho_{\rm spin}({\bf p})=\rho_{\uparrow}({\bf p})-\rho_{\downarrow}({\bf p}).
\end{equation}
The magnetic Compton profile, $J_{\rm mag}(p_{z})$, is defined as the one dimensional (twice integrated) projection
of the electron spin momentum density, $\rho_{\rm spin}({\bf p})$, along the scattering vector which is parallel
(orthogonal) to $p_{z}$ ($p_{x,y}$),
\begin{equation}
J_{\rm mag}(p_{z})=\iint\big[\rho_{\uparrow}({\bf p})-\rho_{\downarrow}({\bf p})\big]~\text{d}p_{x}\text{d}p_{y},
\end{equation}
which is normalized to the electron spin moment, $m^{\rm spin}_{z}$,
\begin{equation}
m^{\rm spin}_{z}=\int J_{\rm mag}(p_{z})~\text{d}p_{z}.
\end{equation}

Magnetic Compton scattering is only sensitive to the spin magnetic moment \cite{cooper:92,carra:96}.
Because only those electrons that contribute to the spin moment of the sample contribute to
the integral of the magnetic Compton profile, it is then possible to determine the spin magnetic
moment, usually by comparison with a measurement, under the same experimental conditions,
of a sample with known spin moment (in this case, fcc Ni).
Since the magnetic Compton profile is the difference between two measured Compton profiles,
components arising from spin-paired electrons cancel, as do most sources of systematic error.

In practice, the spin magnetic moment along the scattering vector is
determined from the so called `flipping ratio', $R$, defined as,
\begin{equation}
R=\frac{I^{\uparrow}-I^{\downarrow}}{I^{\uparrow}+I^{\downarrow}}=\frac{\int J_{\rm mag}(p_{z})~\text{d}p_{z}}{\int J(p_{z})~\text{d}p_{z}}=\frac{m_{z}^{\rm spin}}{N},
\label{flipratio}
\end{equation}
where $I^{\uparrow}$ and $I^{\downarrow}$ are the integrated intensities when the sample moment is aligned
parallel and antiparallel to the (fixed) photon helicity.
By comparing the flipping ratio of the sample in question with that of a calibration sample of known spin
moment, $m^{\rm spin}_{\rm cal}$, the experimental spin moment of the sample in question is then given by,
\begin{equation}
m^{\rm spin}_{\rm exp}=\frac{R_{\rm exp}}{R_{\rm cal}}\frac{N_{\rm exp}}{N_{\rm cal}}m^{\rm spin}_{\rm cal},
\label{flip}
\end{equation}
where $N_{\rm exp}$ and $N_{\rm cal}$ are the number of electrons in the experimental and calibration samples, respectively.
Typically, the chosen calibration sample is fcc Ni with $m^{\rm spin}_{\rm cal}=0.56$~$\mu_{\rm B}$~atom$^{-1}$.

In this study, magnetic Compton profiles were measured along the cubic high symmetry $[100]$, $[110]$ and $[111]$
directions for NiFeCo, NiFeCoCr and NiFeCoCrPd on beamline BL08W at the SPring-8 synchrotron, Japan.
All of the magnetic Compton profiles were recorded at $T=10$~K and $\mu_{0}H_{\rm ext}=\pm2$~T.
All of the samples were cooled in zero field.
The experimental full-width-half-maximum (FWHM) resolution was about $0.45$~a.u. at the Compton peak.
The incident photon energy was $183.4$~keV.
The measured profiles were then corrected for energy dependent detector efficiency, sample absorption,
the relativistic scattering cross-section and multiple scattering.
The data analysis has been described in detail previously \cite{cooper:97,mccarthy:97}.

\subsection{X-ray absorption measurements}

In x-ray absorption spectroscopy (XAS), the x-ray absorption spectrum is given by $\mu(E)=\big[\mu^{+}(E)+\mu^{-}(E)\big]/2$
and the XMCD spectrum is given by $\Delta\mu(E)=\big[\mu^{+}(E)-\mu^{-}(E)\big]$,
where $\mu^{+}(E)$ and $\mu^{-}(E)$ represent the energy dependent absorption cross-sections of soft x-ray photons
with positive and negative helicity, $h^{+}$ and $h^{-}$, respectively.
In practice, the energy is fixed for each data point at a chosen energy, $E$, and the absorption signal is given by,
\begin{equation}
\mu_{E}^{\pm}=\int\mu^{\pm}(E')\delta(E'-E)~\text{d}E',
\label{sxm}
\end{equation}
where the $\delta$-function represents the chosen (Gaussian) energy resolution.

The absorption signals were recorded with the soft x-ray absorption spectrometer on BL25SU \cite{hara:03,hirono:05,nakamura:05,senba:16},
at the SPring-8 synchrotron, Japan, by means of the total electron yield (TEY) method.
This apparatus is equipped with an electromagnet with a maximum applied magnetic field of $\mu_{0}H_{\rm ext}=\pm1.9$~T,
and a cryostat which can cool the sample down to $10$~K.
In this experiment, the energy resolution was set to $E/\Delta E=3000$, where $\Delta E$ is the
Gaussian FWHM.
Soft XMCD spectroscopy using TEY detection is a surface sensitive
magnetic probe which has a probing depth (exponential decay length) of about $1$~nm from the
sample surface \cite{abbate:92,vogel:94,frazer:03}.
For Cr, Mn, Fe, Co and Ni, the spectra were recorded across their respective $L_{2,3}$-edges ($2p\rightarrow3d$
transitions), while for Pd the spectra were recorded across its $M_{2,3}$-edges ($3p\rightarrow4d$ transitions).
All of the spectra were recorded at $T=10$~K and $\mu_{0}H_{\rm ext}=\pm1.9$~T in order to saturate the moments. 
The recorded spectra were normalized by the TEY intensity monitored with a SiC membrane located upstream of the sample.

\subsubsection{Orbital and spin sum rules}

Measurement of the XAS and XMCD spectra permits the determination of the $z$-component (along the x-ray incidence direction) of the orbital and spin magnetic moments
via application of the orbital and spin sum rules \cite{thole:92,carra:93}.
The sum rules are expressed in terms of various integrals over the XAS and XMCD spectra that are given the symbols
$p$, $q$, and $r$ \cite{chen:95}.
For the $3d$ elements,
$p$ is the integral of the XMCD spectrum over only the $L_{3}$-edge, {\it i.e.} the integral ends at the 
onset of the $L_{2}$-edge,
\begin{equation}
p=\int_{L_{3}}\Delta\mu(E)~\text{d}E,
\label{p}
\end{equation}
$q$ is the integral of the XMCD spectrum over both the $L_{2}$ and $L_{3}$ edges,
\begin{equation}
q=\int_{L_{3}+L_{2}}\Delta\mu(E)~\text{d}E,
\label{q}
\end{equation}
and $r$ is the integral of the background corrected XAS spectrum, $\mu^{0}(E)=\mu(E)-f_{\rm bkg}(E)$,
over both the $L_{2}$ and $L_{3}$ edges,
\begin{equation}
r=2\int_{L_{3}+L_{2}}\mu^{0}(E)~\text{d}E,
\label{r}
\end{equation}
where  $f_{\rm bkg}(E)$ is the non-magnetic background which is a quadratic constructed from
the linear gradients of the pre-$L_{3}$- and post-$L_{2}$-edge regions of $\mu(E)$,
and two $\arctan$ step functions (one centered at the $L_{2}$-edge and one centered at the $L_{3}$-edge)
for the continuum absorption \cite{goering:05}.

For the $3d$ elements, the sum rules then state that the $z$-component of the orbital magnetic moment
(in units of $[\mu_{\rm B}$ atom$^{-1}]$) is given by, 
\begin{equation}
m^{\rm orb}_{z}=\frac{-4q(10-n_{3d})}{3rP_{c}\cos(\alpha)},
\label{orb}
\end{equation}
and the $z$-component of the effective spin magnetic moment (in units of $[\mu_{\rm B}$ atom$^{-1}]$) is given by, 
\begin{equation}
m^{\rm spin,eff}_{z}=m^{\rm spin}_{z}-7\langle T_{z}\rangle=\frac{-(6p-4q)(10-n_{3d})}{rP_{c}\cos(\alpha)},
\label{spin}
\end{equation}
where $m^{\rm spin}_{z}=-2\langle S_{z}\rangle$ (in Hartree atomic units),
$\langle S_{z}\rangle$ and $\langle T_{z}\rangle$ are the expectation values of the 
$z$-components of the spin angular momentum operator and magnetic dipole operator, respectively,
$n_{3d}$ is the site averaged number of $3d$ electrons of the element in question (determined by electronic
structure calculations), $P_{c}=0.96$ \cite{hirono:05} is the degree of circular polarization, and $\alpha=10^{\circ}$ is the
angle between the applied field direction and the incident x-ray direction \cite{nakamura:05}.
Note that $\langle T_{z}\rangle$ is generally assumed to be negligible for atoms in a cubic environment
\cite{carra:93,obrien:94}.
For all of the measured alloys, the spin sum rule moments of Cr were doubled because, for Cr, the spin–orbit splitting of the $2p_{3/2}$ and $2p_{1/2}$ core levels is small
resulting in significant overlap of the $L_{3}$ and $L_{2}$ absorption edges and, hence, spin sum rule moments that are too small by about a factor of two \cite{goering:05}.
Accordingly, the spin sum rule moments are expected to be more accurate for Fe, Co and Ni because of the larger spin-orbit splitting.

\subsubsection{Element specific hysteresis loops}

The variation of the total magnetic moment, $m^{\rm tot}_{z}=m^{\rm orb}_{z}+m^{\rm spin}_{z}$,
with $H_{\rm ext}$ is proportional to,
\begin{equation}
m^{\rm tot}_{z}(H_{\rm ext})\propto\frac{\Delta\mu_{L_{2}}(H_{\rm ext})}{\mu_{L_{2}}(H_{\rm ext})}-\frac{\Delta\mu_{L_{3}}(H_{\rm ext})}{\mu_{L_{3}}(H_{\rm ext})},
\label{MH}
\end{equation}
where the on-edge XMCD signals are normalized by their respective XAS signals in order to account for the non-linear variation
of the TEY signal with $H_{\rm ext}$ \cite{goering:00} (note that for Pd, we replace $L_{2,3}$ with $M_{2,3}$).
All of the hysteresis loops were recorded at $T=10$~K between applied magnetic fields of $\mu_{0}H_{\rm ext}=\pm1.9$~T.
The measured element specific hysteresis loops for Cr, Mn, Fe, Co and Ni were scaled to equal their respective
$m^{\rm tot}_{z}$ as determined by the sum rule analysis of the recorded spectra, while that of Pd was scaled
to the total Pd $d$-electron magnetic moment given by KKR-CPA calculations.

\subsection{Electronic structure calculations}

The KKR method \cite{korringa:47,kohn:54,ebert:11} was used to calculate the electronic structure
and the CPA \cite{soven:67,gyorffy:72,faulkner:80} was used to treat the compositional disorder.
The KKR-CPA calculations were performed with the Munich SPR-KKR code \cite{sprkkr} within the atomic sphere approximation (ASA).
The core configuration for all elements except Pd was $1s^{2}2s^{2}2p^{6}3s^{2}3p^{6}$,
and the core configuration for Pd was $1s^{2}2s^{2}2p^{6}3s^{2}3p^{6}3d^{10}4s^{2}4p^{6}$.
The lattice constants for NiFeCo, NiFeCoCr and NiFeCoCrMn were $3.577$~\AA, while that of NiFeCoCrPd was $3.657$~\AA~due to the atomic size mismatch \cite{lucas:11,dahlborg:16}.
For NiFeCo, NiFeCoCr and NiFeCoCrMn, the muffin-tin radii were $2.38$~a.u. while for NiFeCoCrPd the muffin-tin radii were $2.44$~a.u. and $1200$ $k$-points were used to sample the irreducible wedge of the first Brillouin zone.
The local density approximation (LDA) to the exchange-correlation energy functional was
Vosko-Wilk-Nusair (VWN) \cite{vosko:80}.

From the converged electronic structures, directional magnetic Compton profiles were calculated along the same directions
as measured in the experiment.
For comparison with the experimental magnetic Compton profiles, all of the calculated magnetic Compton profiles were
convoluted with a Gaussian function with the FWHM equal to the experimental resolution at the Compton peak ($0.45$~a.u.).

\section{Results}

\begin{figure*}[t!]
\centerline{\includegraphics[width=1.0\linewidth]{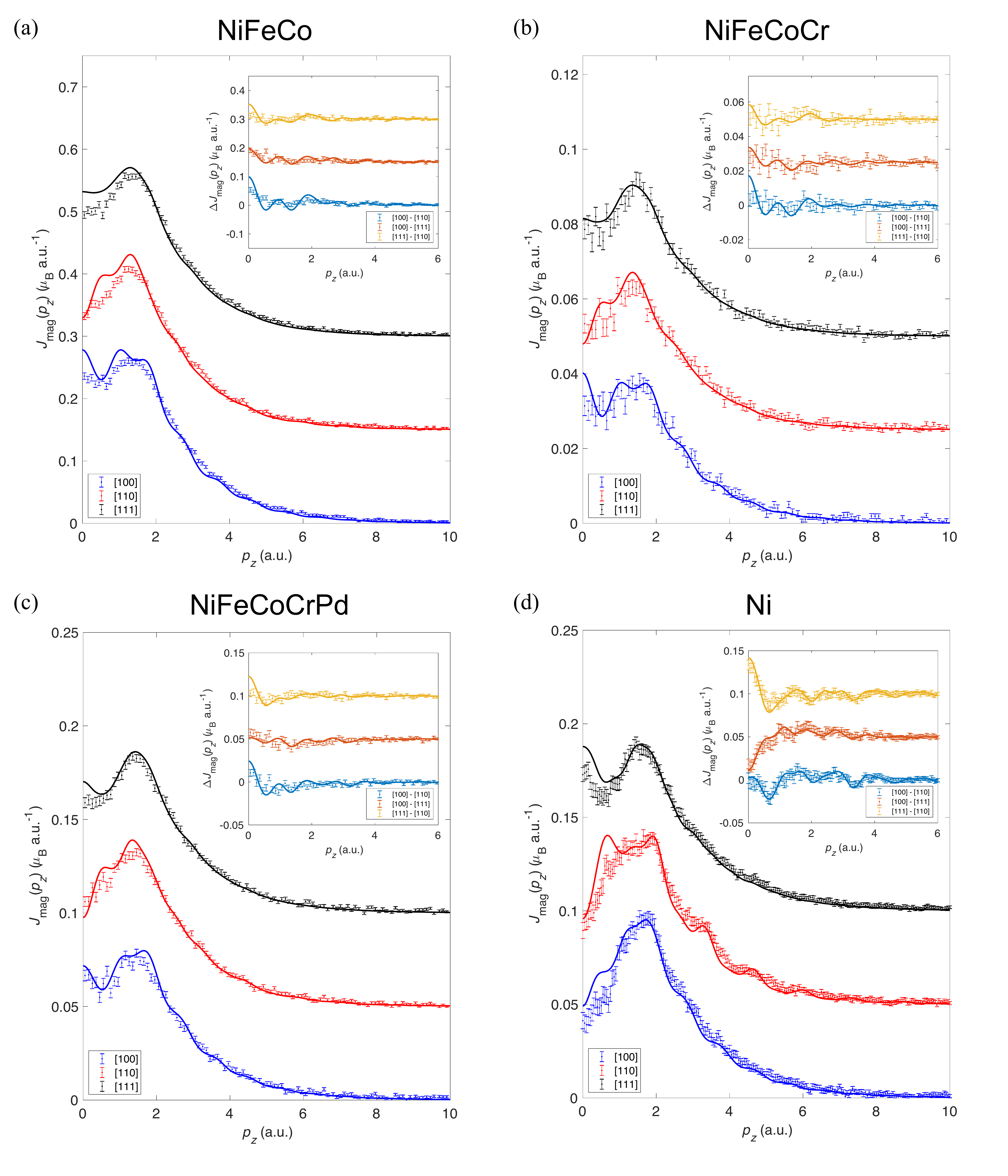}}
\caption{
Experimental (points) and calculated (lines) magnetic Compton profiles, $J_{\rm mag}(p_{z})$,
of (a) NiFeCo, (b) NiFeCoCr, (c) NiFeCoCrPd and (d) fcc Ni recorded with the scattering vector parallel to the
$[100]$ (blue), $[110]$ (red) and $[111]$ (black) high symmetry crystallographic directions.
The insets show the directional differences of the magnetic Compton profiles, $\Delta J_{\rm mag}(p_{z})$,
between the $[100]$ and $[110]$ (light blue), $[100]$ and $[111]$ (orange), and $[111]$ and $[110]$ (yellow) directions.
The directional profiles in each panel and directional differences in each inset have been offset by steps of
(a) $0.15$, (b) $0.025$, (c) $0.05$ and (d) 0.05~$\mu_{\rm B}\,{\rm a.u.}^{-1}$ for clarity.
The experimental and calculated directional profiles have been normalized to their experimentally determined
bulk spin moments (from Eq.~\ref{flip}) along their respective directions.
The error bars are statistical errors of one standard deviation.
Note that the error bars are larger when the bulk spin magnetic moment is smaller
due to the measured signal being proportional to the spin magnetization. 
The fcc Ni experimental profiles are taken from Ref.~\cite{dixon:98}.
}
\label{compton}
\end{figure*}

\begin{figure*}[t!]
\centerline{\includegraphics[width=1.0\linewidth]{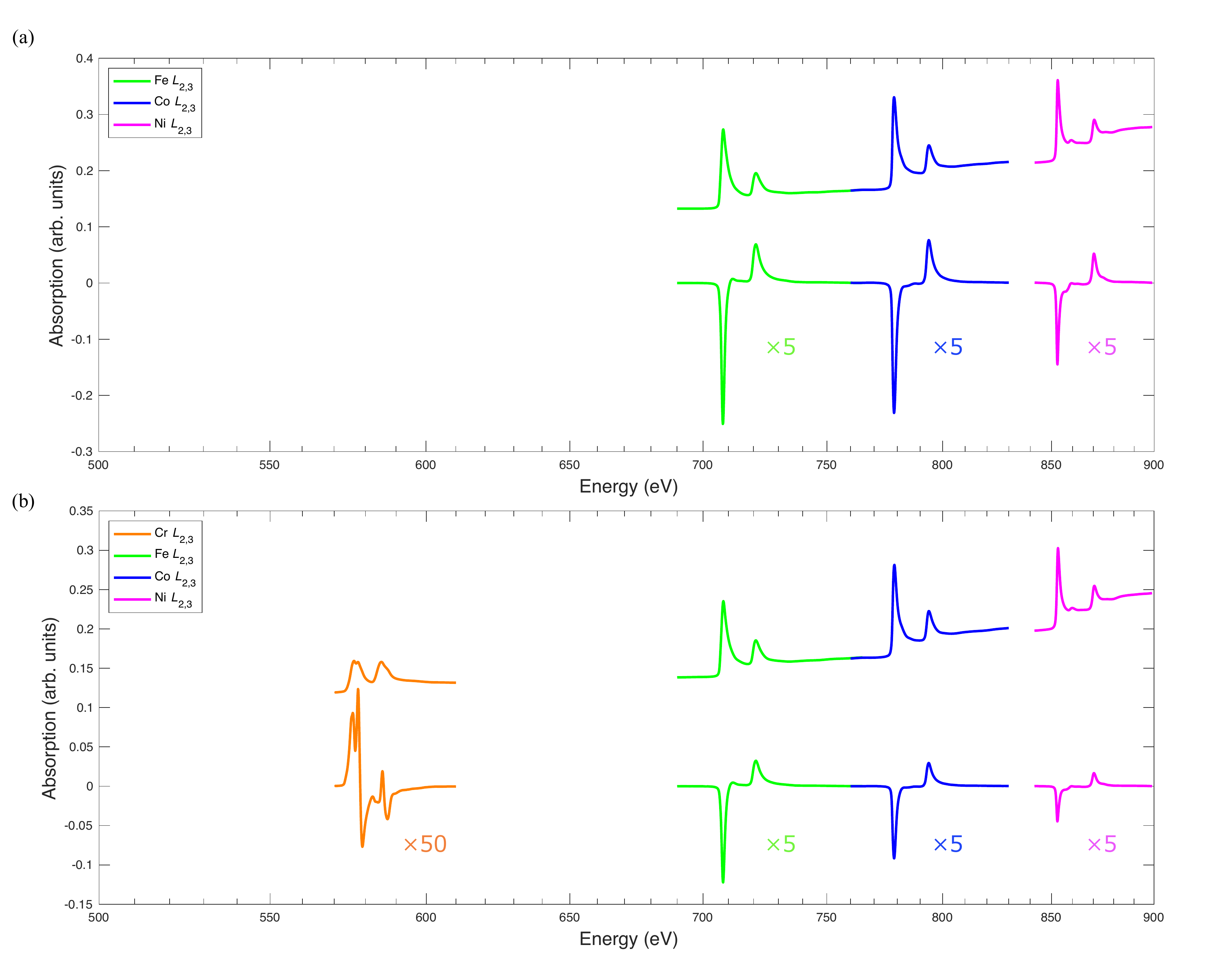}}
\caption{XAS (top) and XMCD (bottom) spectra of Ni, Fe, Co and Cr in (a) NiFeCo and (b) NiFeCoCr.
The multiplication labels indicate the scaling of the XMCD signals.
All of the XAS signals have been offset by a constant value of $0.1$ for clarity.
Note that the incident photon energy is plotted on a logarithmic scale (the $50$~eV energy intervals get closer together with increasing energy) for clarity because at lower energies
the core-level spin-orbit splitting of species with lower atomic number is smaller.}
\label{3-4_spec}
\end{figure*}

\begin{figure*}[t!]
\centerline{\includegraphics[width=1.0\linewidth]{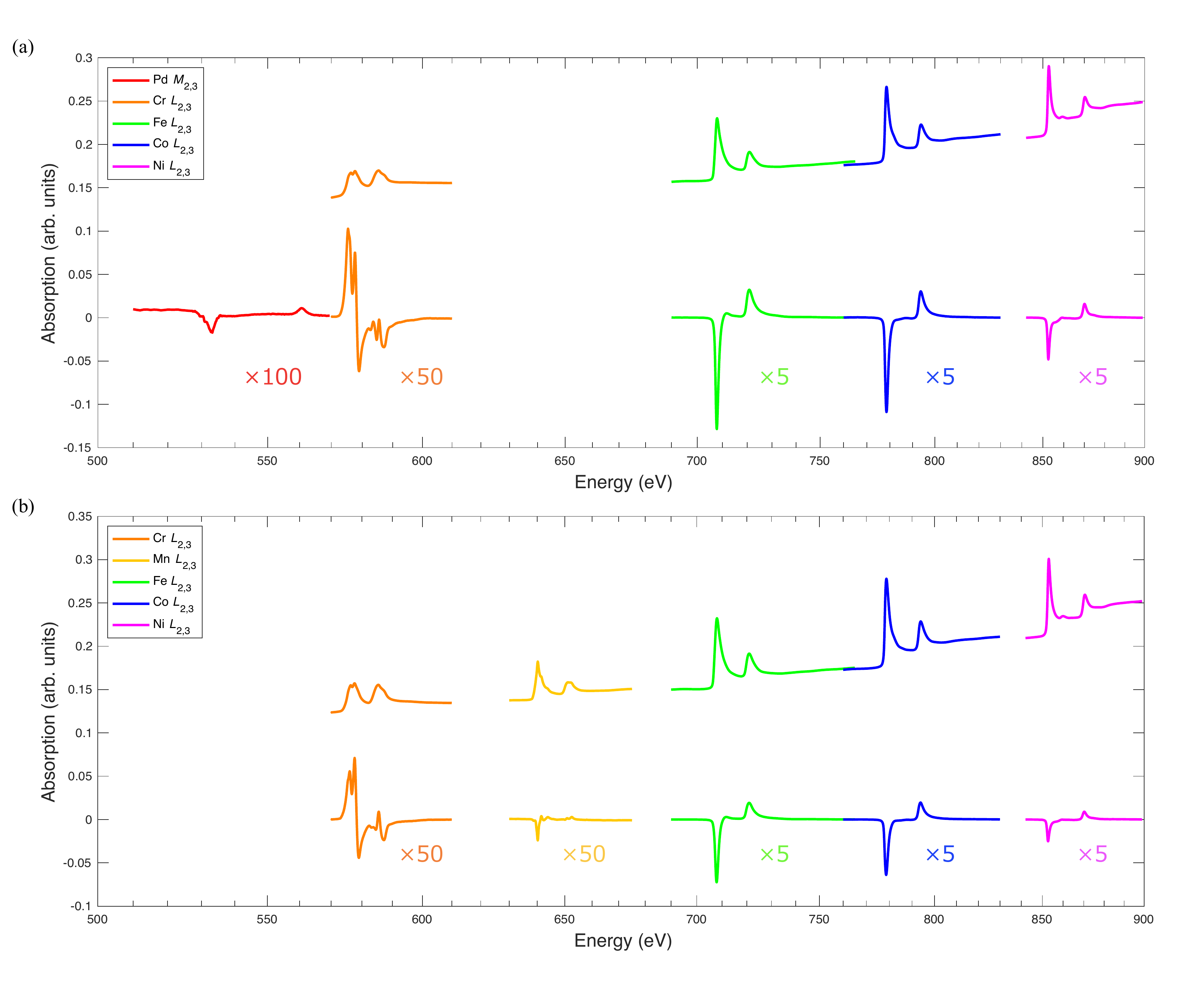}}
\caption{XAS (top) and XMCD (bottom) spectra of Ni, Fe, Co, Cr, Pd and Mn in (a) NiFeCoCrPd and (b) NiFeCoCrMn.
The multiplication labels indicate the scaling of the XMCD signals.
All of the XAS signals have been offset by a constant value of $0.1$ for clarity.
Note that the incident photon energy is plotted on a logarithmic scale (the $50$~eV energy intervals get closer together with increasing energy) for clarity because at lower energies
the core-level spin-orbit splitting of species with lower atomic number is smaller.}
\label{HEA_spec}
\end{figure*}

\begin{figure*}[t!]
\centerline{\includegraphics[width=1.0\linewidth]{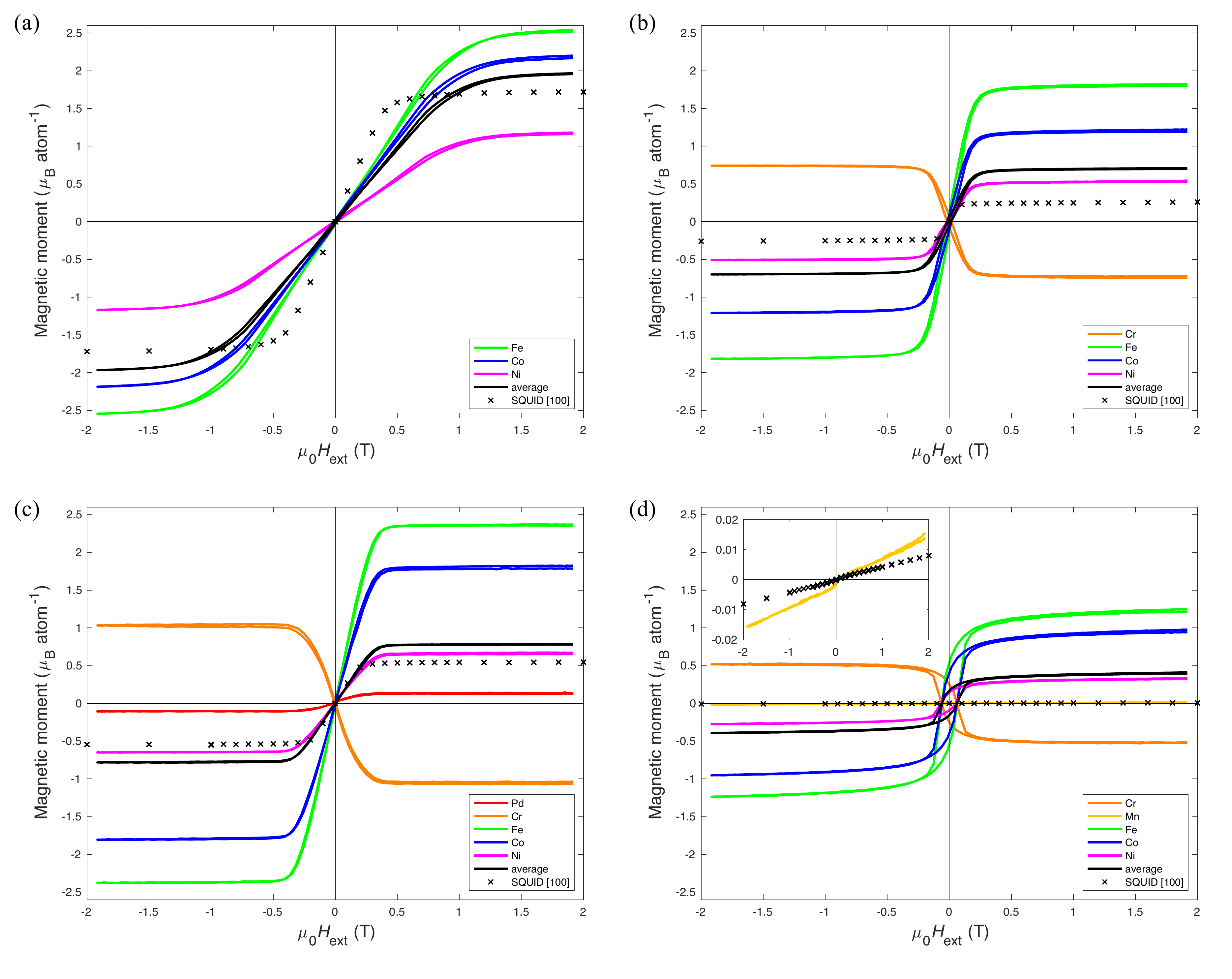}}
\caption{Element specific magnetization curves for (a) NiFeCo, (b) NiFeCoCr, (c) NiFeCoCrPd and (d) NiFeCoCrMn determined by the XMCD measurements together with their respective magnetization curves determined by the SQUID measurements.
The saturated magnetic moments for each atomic species have been scaled to the corresponding total moments determined by the orbital and spin
sum rules (except for Pd where it was scaled to the Pd $d$-electron moments from the KKR-CPA calculations).
The inset of (d) shows a close up of the NiFeCoCrMn magnetization curve from the SQUID measurements together with the Mn magnetization curve from the XMCD measurements.
The inset has the same axis labels and units as those of the main panel.}
\label{MH_plots}
\end{figure*}

The experimental and calculated MCPs of NiFeCo, NiFeCoCr, NiFeCoCrPd and Ni
(the Ni experimental data were previously reported in Ref.~\cite{dixon:98}) are shown in Fig.~\ref{compton}(a), (b), (c) and (d),
respectively, resolved along the cubic $[100]$, $[110]$ and $[111]$ high symmetry crystallographic directions. The inset to each
figure shows the anisotropy of the spin density in momentum space in the differences between MCPs measured
along different crystallographic directions.
The areas under the experimental profiles are equal to the experimental bulk spin moments (determined by Eq.~\ref{flip}) and are listed in Table~\ref{tab1}.
Superficially, the general shape of the MCPs of the Cantor-Wu alloys are
similar to those of Ni; they are finite at $p_{z}=0$~a.u. and rise to a broad maximum
around $1<|p_{z}|<2$~a.u. with a tail that asymptotically approaches zero with
increasing momentum.
The reduced intensity at low momentum in an MCP can be due to a number of factors. It is at low momentum that the most itinerant valence electrons are contributing to the momentum density. The rapid increase in intensity in a MCP could come from the negative net spin polarization of the delocalized
(hence relatively localized in momentum space) $sp$-electrons that screen the more localized (hence delocalized in momentum space)
$d$-electron moment (which has a much larger positive net spin polarization).
Note that the degree of negative polarization of the $sp$-electrons in metallic $3d$-moment
systems is, typically, underestimated by LDA density functional theory (DFT) calculations
\cite{kubo:90,timms:90,sundararajan:91,dixon:98}.
It could also be due to the radial behavior of the 
wavefunction in momentum space which for a $d$-electron orbital goes like $p_{z}^{2}$ \cite{mijnarends:73}, resulting in a small contribution to the
momentum density at low momentum which grows strongly as $p_{z}^{4}$. Finally, it could be a consequence of the Fermi surface, 
for example through the presence of a majority hole-pocket or minority electron pocket at the $\Gamma$ point, but this
is not the case in these alloys. 
The disagreement at low momentum shows that the negative spin polarisation is sensitive to the treatment of exchange and correlation \cite{dixon:98}.
Dynamical mean field theory has been investigated as a possible solution to this problem \cite{benea:12,chioncel:14a,chioncel:14b}.
 
On closer inspection, however, it is clear that the MCPs of Ni show much
more structure and anisotropy than those of the Cantor-Wu alloys, particularly for $|p_{z}|>2$~a.u. where
higher momentum Umklapp features are much more prominent. These features are due to the Fermi surface and arise
from majority and minority bands crossing the Fermi energy at different crystal momenta (and different 
real momenta) leading to small peaks or troughs (for example, see the theoretical profiles in Ref.~\cite{dixon:98}). Although
these sharp structures are smeared by the typical experimental resolution (the FWHM is about half
of the size of the Brillouin zone), features such as shoulders ({\it e.g.} Ni $[110]$ at $|p_{z}|\approx3.5$~a.u.) can be resolved in the Ni data (Fig.~\ref{compton} (d)).
The MCPs of each Cantor-Wu alloy are remarkably similar along each crystallographic direction and,
for a given direction, the profile shapes are quite similar between
the different Cantor-Wu alloys.
In comparison to Ni, the MCPs of the Cantor-Wu alloys appear smeared.
Robarts {\it et al.} very recently used high resolution (non-magnetic) Compton scattering
to experimentally determine the bulk Fermi surface geometry of NiFeCoCr, and found that it is smeared
over $\sim40\%$ of the Brillouin zone \cite{robarts:20}. Such smearing implies a short electron mean-free-path and thus a high
residual resistivity. The Bloch spectral functions calculated by Mu {\it et al.} \cite{mu:19} 
show that the majority spin Fermi surface remains very sharp in NiFeCo (since the majority spin band centers align with each other for these
$3d$ elements), explaining the relatively low resistivity as the conductivity
short circuits via the majority spin channel. 
Focusing on the insets in Fig.~\ref{compton} (a)---(d), the intensities of the anisotropies are markedly smaller than the calculations predict
for all the Cantor-Wu alloys. For Ni (Fig.~\ref{compton}(d)), the agreement between the experiment and calculation is excellent. This implies that an inadequate DFT description (due to the use of Kohn-Sham wavefunctions \cite{lam:74} and/or inadequate exchange-correlation potential) 
is unlikely to be responsible for the overestimation of the anisotropy in the calculations for the Cantor-Wu alloys. Furthermore, it indicates 
whatever inadequacies there are in the calculation for Ni, they disappear in the double difference between spins and 
crystallographic directions. This anisotropy is not observably smaller in NiFeCo, where calculations suggest that only the minority Fermi surface is smeared \cite{mu:19}. 
The apparent `isotropy' of the MCPs of the Cantor-Wu alloys is, in fact,
symptomatic of the smearing of the electronic structure by the compositional disorder. The high residual 
resistivity of Cr-containing alloys, being emblematic of a strongly smeared Fermi surface, suggests that the electrons are going to
be ignorant of any phenomena involving coordination over distances much longer than their mean-free-path. Given that this distance
is of the order of the lattice spacing, the lack of anisotropy in the momentum space spin density is perhaps not surprising.

The XAS and XMCD spectra of the measured absorption edges of NiFeCo and NiFeCoCr are shown in Fig.~\ref{3-4_spec},
and those of NiFeCoCrPd and NiFeCoCrMn are shown in Fig.~\ref{HEA_spec}.
Note that the Pd XAS spectrum is not shown because the SiC membrane (whose TEY signal is used to normalize the measured
spectra to the incident photon flux) has some adsorbed oxygen from exposure to air and the Pd $M_{2,3}$-edges are in the
same energy range as the O $K$-edge meaning that quantitative Pd moments cannot be determined.
In all of the measured alloys, the Ni, Fe and Co XAS and XMCD spectra have very similar shapes to that of their respective pure metals
\cite{chen:95,nakajima:99} indicating undetectable levels of surface oxidation of these elements
(the XAS of Ni-, Fe- and Co-oxides exhibit split peak structures at both the $L_{2}$- and $L_{3}$-edges due to multiplet effects,
see {\it e.g.} \cite{crocombette:95,alvarenga:01,alvarenga:02}).
The relative sizes of the jumps in the XAS at the $L_{3}$-edge ($\mu_{L_{3}}-\mu_{\text{pre-}L_{3}}$)
of Ni, Fe and Co (and Cr in the Cr-containing alloys) within an alloy are the same between different alloys indicating
that the relative chemical concentrations of these elements are not changing between the different alloys
($\mathrm{Ni}:\mathrm{Fe}:\mathrm{Co}\ (:\mathrm{Cr})$ is constant).

Non-zero XMCD signals are observed for every element in every alloy indicating finite magnetic moments,
although for each element the sizes of the XMCD signals relative to their respective XAS signals vary
significantly between alloys.
Except for Cr, the XMCD signals of all of the measured elements of each alloy are negative at the
$L_{3}$-edge and positive at the $L_{2}$-edge (negative at the $M_{3}$-edge and positive at the $M_{2}$-edge for Pd)
indicating that the $z$-components of the site-averaged moments are parallel to the applied magnetic
field and are ferromagnetically coupled to each other.
In each Cr-containing alloy, the Cr XMCD signal is positive at the $L_{3}$-edge and negative at
the $L_{2}$-edge indicating that the $z$-component of the site-averaged moment is unambiguously aligned antiparallel
to the applied magnetic field and that Cr is therefore antiferromagnetically coupled to the other elements
in agreement with first-principles calculations \cite{jin:16,schneeweiss:17,mu:19,mu:19a,lagosphd}.
For a quantitative determination of the element specific orbital and spin magnetic moments,
the orbital and spin sum rules (Eqs.~\ref{orb} and \ref{spin}, respectively) were applied to
the measured spectra of Ni, Fe, Co, Cr and Mn.
The values obtained are are listed in Table~\ref{tab1}.

The $m_{z}^{\rm tot}(H_{\rm ext})$ curves determined by Eq.~\ref{MH} from the XAS and XMCD data
are shown in Fig.~\ref{MH_plots}, together with the bulk total moment curves obtained from the SQUID
magnetometry measurements.
NiFeCo has the largest moment, so would be most susceptible to demagnetization effects,
although it is unclear how to correct for demagnetization effects at the sample surface.
Nevertheless, the magnetic moments of all alloys are saturated at the maximum applied magnetic field of
$\mu_{0}H_{\rm ext}=\pm1.9$~T, which is where the XAS and XMCD spectra were recorded.
The element specific magnetization curves of the spin glass alloys NiFeCoCr and NiFeCoCrMn exhibit hysteresis
which is related to the energy barrier that must be overcome in order for the frozen magnetic moments to rearrange
themselves.

\section{Discussion}

\begin{figure}[t!]
\centerline{\includegraphics[width=1.0\linewidth]{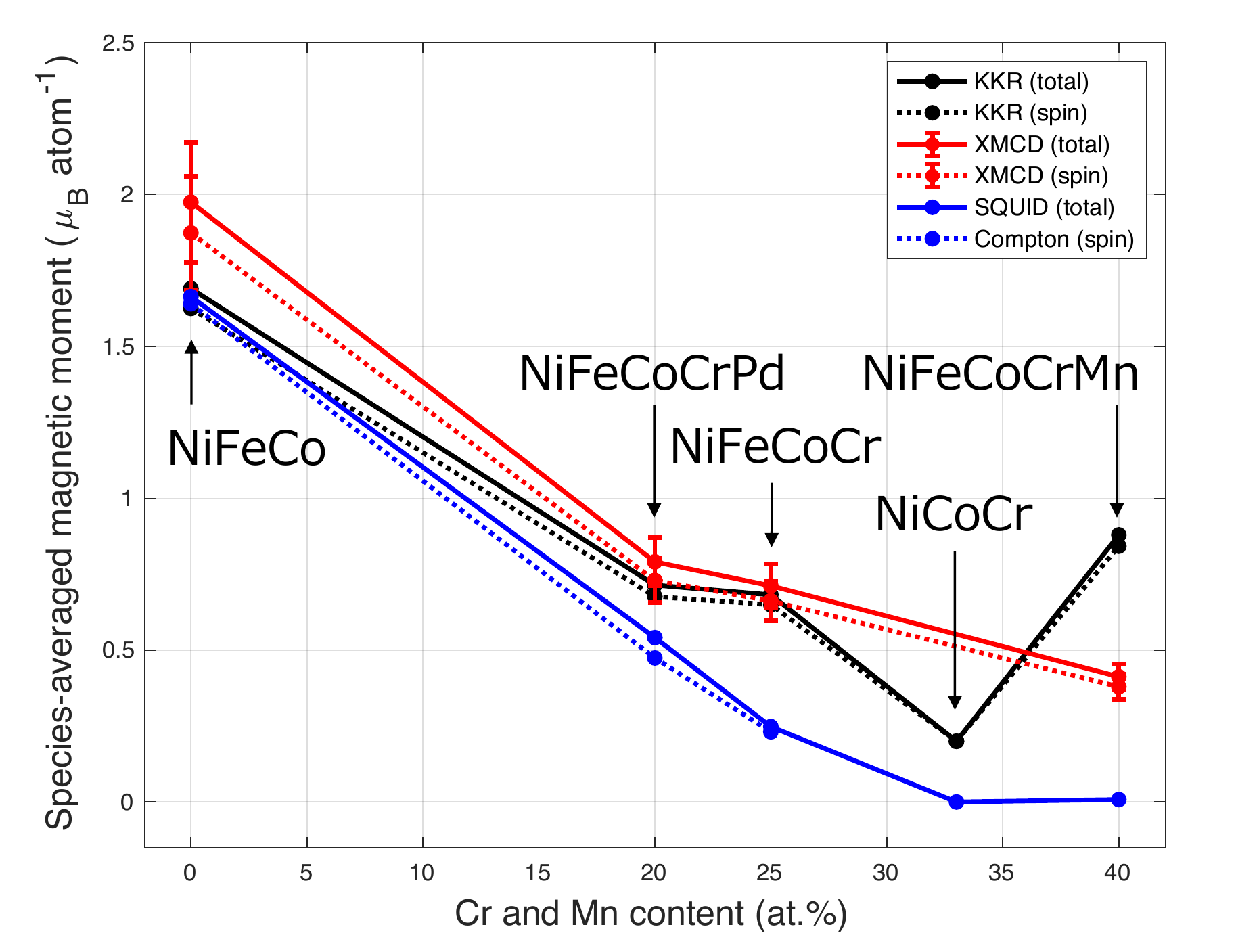}}
\caption{
Species averaged total moments and spin moments as a function of the relative combined Cr and Mn concentration
for NiFeCo ($0$~at.\% Cr+Mn), NiFeCoCr ($25$~at.\% Cr+Mn), NiFeCoCrPd ($20$~at.\% Cr+Mn) and NiFeCoCrMn ($40$~at.\% Cr+Mn)
as determined by KKR-CPA calculations (black), the XMCD orbital and spin sum rules (red),
and bulk (SQUID and Compton) measurements (blue).
The experimental (SQUID) and calculated (KKR) total moment of NiCoCr ($33$~at.\% Cr+Mn)
from Ref.~\cite{sales:16} are also plotted.
Solid and dotted lines are guides to the eye for the total moment and the spin moment, respectively.
The orbital moment is given by the difference between the total moment and the spin moment.
The orbital moment of NiCoCr has not been reported so we have set the calculated $m^{\rm tot}=m^{\rm spin}$,
and magnetic Compton scattering was not performed on NiCoCr or NiFeCoCrMn so is not plotted.
The error bars are statistical errors of one standard deviation.
The error bars from the bulk measurements are approximately equal to the size of the points.
}
\label{HEA_mom}
\end{figure}

All of the experimental and calculated magnetic moments determined in this study are summarized in Table~\ref{tab1}.
In all of the alloys measured here, both the sum rules and KKR-CPA calculations predict positive values for
the orbital and spin moments of Ni, Fe, Co and Mn, while the orbital moment is positive and the spin moment
is negative for Cr (antiparallel spin and orbital moments are expected in Cr from Hund's rules).
For every element in every alloy, the orbital moments are essentially quenched due to the (approximately) cubic symmetry;
Co has the largest orbital sum rule moment, but it is never more than about $10\%$ of the spin sum rule moment. Eriksson {\it et al.} predicted a large orbital moment for fcc Co \cite{eriksson:90}.

In NiFeCo, the experimentally determined bulk moments agree remarkably well with the KKR-CPA calculations.
Given that NiFeCo is the least compositionally complex of the alloys investigated here and that all of the
elements present have ferromagnetic couplings, this is hardly surprising.
For NiFeCo, NiFeCoCr and NiFeCoCrPd, the Fe moments from the sum rules also agree remarkably well with the KKR-CPA calculations.
However, the Ni and Co orbital and spin sum rule moments are significantly overestimated leading to a species averaged
moment that is larger than both the KKR-CPA calculations and bulk measurements.
In fact, compared to the KKR-CPA calculations the Ni spin sum rule moment is systematically overestimated by $30$-$50\%$ 
in NiFeCo, NiFeCoCr and NiFeCoCrPd, while the Co spin sum rule moment is systematically overestimated by $10$-$20\%$
in NiFeCo and NiFeCoCrPd.
In NiFeCoCrPd, the sum rule moments show reasonable agreement with the average KKR-CPA moments,
but both are significantly larger than the bulk measurements.

The situation is much more complicated for NiFeCoCr and NiFeCoCrMn, both of which have been reported to exhibit spin glass behavior  \cite{kao:11,jin:16,schneeweiss:17}.
In NiFeCoCr, the element specific sum rule and calculated moments are in reasonable agreement
with each other, but the species averaged moments are much larger than those determined by the
SQUID and Compton measurements.
It is worth noting that Jin {\it et al.} found a similar discrepancy between their experimental
($m^{\rm sat}=0.24$~$\mu_{\rm B}$) and calculated ($m^{\rm tot}=0.66$~$\mu_{\rm B}$~atom$^{-1}$)
species averaged total magnetic moment of NiFeCoCr to the one found here, which they suggested may
well be indicative of a more complex, non-collinear, magnetic ground state than allowed by their
KKR-CPA calculations, which were restricted to collinearity \cite{jin:16}.
In NiFeCoCrMn, the element specific sum rule and calculated moments are in disagreement
with each other, and the species averaged moments are much larger than those determined by our SQUID measurements and those of Jin {\it et al.} \cite{jin:16}.
In terms of the elemental moments, the largest discrepancy is seen for Mn; the KKR-CPA calculation predicts
a large spin moment of $1.8$~$\mu_{\rm B}$~atom$^{-1}$, while the observed XMCD signal is extremely small giving a
spin sum rule moment less than $0.01$~$\mu_{\rm B}$~atom$^{-1}$.
Again, this suggests that the KKR-CPA calculations employed here are not sophisticated enough to describe
the real magnetic state of these two alloys.
Indeed, Schneeweiss {\it et al.} \cite{schneeweiss:17} and Mu {\it et al.} \cite{mu:19} independently performed more sophisticated spin collinear supercell calculations and
disordered local moment (DLM) KKR-CPA calculations, respectively, which both predict that in NiFeCoCrMn there are approximately equal
populations of Mn atoms with large moments ($1.5$-$2$~$\mu_{\rm B}$~atom$^{-1}$) aligned parallel (Mn$^{\uparrow}$)
and antiparallel (Mn$^{\downarrow}$) to the spin quantization axis which would give us a species averaged moment of
$0.5$~$\mu_{\rm B}$~atom$^{-1}$ and a vanishing Mn site averaged moment which is in much better agreement with the
XMCD measurements.
Interestingly, Mu {\it et al.} \cite{mu:19a} also found an unconventional CPA ground state in NiCoMn
which distinguishes two equally populated Mn CPA components with large but oppositely oriented spin moments and,
using spin spiral calculations, they further demonstrated this calculated ground state
is most energetically favorable in the presence of spin non-collinearity.
The present XMCD measurements on NiFeCoCrMn provide the first strong experimental evidence for the existence of such a bizarre magnetic state,
the prediction of which could be considered a triumph of the KKR-CPA-DLM theory.

In order to compare with the experimental MCPs, the calculated MCPs were normalized to the experimental spin moment.
This procedure is not strictly valid given that the calculated moments (listed in Table~\ref{tab1})
are related to the exchange splitting of the majority and minority spin bands, which could change the
Fermi surface size and topology, to both of which the shape of the magnetic Compton profile is sensitive \cite{dixon:98}.
Nevertheless, the calculations normalized to the experimental bulk spin moment provide
an excellent description of the experimental data.
It is well known that the LDA is a mean-field (Stoner) level theory which neglects spin fluctuations that can renormalize the magnetic moment \cite{aguayo:04}.
In fact, it is not unusual to have the calculations predicting larger magnetic moments than experiment and previous Compton scattering experiments \cite{haynes:12} have required a similar rescaling of the profile area in order to agree with the experimental data, with the factor as large as three in UCoGe \cite{butchers:15}.

The species averaged total and spin moments from the different experimental techniques and calculations
are shown in Fig.~\ref{HEA_mom} as a function of the concentration of antiferromagnetically coupled
elements (Cr+Mn) in each alloy studied here, together with those of NiCoCr from Ref.~\cite{sales:16}.
The total and spin moments from the bulk measurements (SQUID and Compton, respectively)
decrease approximately linearly with increasing (Cr+Mn) concentration.
Compared with the bulk measurements, the sum rules systematically overestimate the average spin and,
therefore, average total moments of each alloy, whereas the average orbital moments agree within the error bar.
There may be significant errors in the spin sum rule moments \cite{piamonteze:09}.
The orbital and spin sum rules were originally formulated on the basis of a single ion in cubic symmetry, which
has a well defined number of $d$-electrons.
First-principles calculations indicate that $\langle T_{z}\rangle$ can
reach values of $8.5\%$, $12\%$, and $15\%$ of $\langle S_{z}\rangle$
at the Fermi level for the Fe(001), Ni(001), and Co(0001) surfaces, respectively \cite{wu:94,wu:94a}.
It is worth noting that, through fits of the XAS spectra, Goering \cite{goering:05} determined that the spin sum rule moments
of metallic Fe, Co and Ni are overestimated by about $25\%$, $35\%$ and $45\%$, respectively.
Furthermore, calculations show that electron bands become flatter (narrower bandwidth) at the surface compared
with the bulk due to the loss of near neighbors (which reduces the hopping integral) leading to enhanced spin
and orbital magnetic moments at the surface \cite{eriksson:91,eriksson:92,keshavarz:15}.

At this point, it is worth considering that the KKR-CPA calculations employed here might actually
be representative of the surface magnetic state.
Indeed, Fig.~\ref{HEA_mom} shows that the sum rule and KKR-CPA species averaged moments show qualitative
agreement over the whole series of alloys studied here, except for NiFeCoCrMn which can also be brought
into agreement through more sophisticated calculations \cite{schneeweiss:17,mu:19}.
It is clear that the internal magnetic structures of the Cr and Mn-containing alloys are much more
complicated than the surface sensitive measurements suggest, and might be subject to strong spin
fluctuations that suppress the bulk moments \cite{sales:16}.
The differences between the experimental bulk and theoretical KKR-CPA moments might be due to short range compositional order in these alloys 
\cite{lucas:12,zhang:14,tamm:15,pickering:16,miracle:17,zhang:17,dong:19,zhang:20},
which is not considered in the single-site version of the KKR-CPA calculations used in this study (cluster versions of the CPA exist which can treat short-range compositional order, but such calculations are beyond the scope of this study).
For example, first-principles calculations indicate that structural ordering of Cr in NiFeCoCr
(with Cr atoms located on the corners of the cubic unit cell and randomized Ni, Fe and Co atoms, {\it i.e.} the L$1_{2}$ structure)
can relieve the frustrated magnetic interactions leading to a lower bulk total moment (due to a larger antiparallel Cr moment) \cite{niu:15}.
Note that MCPs determined from supercell calculations of NiFeCoCr in the L$1_{2}$ structure show much worse agreement with experiment.

Finally, although the alloys have been chemically etched and Ar ion sputtered to remove the surface oxide layer,
which is evident as the Ni, Fe and Co $L_{2,3}$-edges do not appear to have significant oxide contributions in
their spectra, the complete removal of Cr-oxide is difficult.
It is also worth considering the thermodynamic stability of transition metal oxides, as represented graphically in the Ellingham diagram \cite{ellingham:44} which plots the change in the Gibbs free energy as a function of temperature for the formation of various transition metal oxides from their respective pure metals.
The change in the Gibbs free energy of formation for Cr$_{2}$O$_{3}$ and MnO is much more negative than for
NiO, CoO, and the Fe-oxides (FeO, Fe$_{3}$O$_{4}$ and Fe$_{2}$O$_{3}$) meaning that Cr and Mn will always be
preferentially oxidized by any residual oxygen in the UHV chamber.
In all of the Cr-containing alloys, the Cr XAS spectra have split peak structures at the $L_{3}$-edge.
This could be due to contributions of both metallic Cr \cite{tomaz:97} and Cr-oxides \cite{gaudry:06}.
If present, Cr$_2$O$_3$ is only likely to exist very near to the surface and would naturally provide
the observed corrosion resistance \cite{qiu:17}.

\section{Conclusion}

Magnetic field dependent synchrotron x-ray experiments with circularly polarized photons
and bulk magnetometry measurements were performed on a set of medium (NiFeCo and NiFeCoCr)
and high (NiFeCoCrPd and NiFeCoCrMn) entropy Cantor-Wu alloys.
The bulk spin momentum densities probed by magnetic Compton scattering are remarkably isotropic and this is a consequence of the smearing of the
electronic structure by the compositional disorder.
The bulk spin moments are in good agreement with the total moments from bulk magnetometry measurements indicating that the orbital moments are essentially quenched due to the (approximately) cubic symmetry.
Finite XMCD signals were recorded for every element in every alloy indicating differences in the populations of the majority and
minority spin states (implying finite magnetic moments) and revealed that the Cr spin moments in the Cr-containing alloys
are unambiguously aligned antiparallel to the bulk total moment.
In NiFeCoCrMn, the total Mn magnetic moment is almost zero which suggests from previous work
that this may be due to an approximately equal number of measured Mn moments which are
parallel and antiparallel to the external field.
Significant discrepancies between the experimental bulk and surface moments have been observed,
and these are not in complete agreement with many of the KKR-CPA calculated moments.
There could be contributions from short range ordering in these samples or more complex alignment of the moments,
which the calculations do not consider.
From this study, a picture of the magnetism of the Cantor-Wu alloys emerges in which their bulk magnetic moments are increasingly suppressed with increasing concentration of antiferromagnetically coupled elements in the solid solution while the surface magnetic moments remain largely oblivious to these suppression mechanisms.

Looking forwards, questions still remain about the nature of the (apparent lack of) magnetism in NiCoCr.
In KKR-CPA calculations, the magnetism of NiCoCr$_{x}$ decreases linearly with increasing Cr content as the quantum critical point ($x\approx1$) is approached,
but decreases exponentially in bulk magnetometry measurements \cite{sales:16}.
Therefore, it would be interesting to repeat the XMCD measurements on NiCoCr$_{x}$ with various compositions
encompassing the quantum critical point to determine whether there are detectable magnetic moments at the sample surface that
are suppressed in the bulk, and to understand their variation with Cr content compared with the bulk moments either side of the quantum critical point.
Another avenue worth exploring would be to investigate the spin dynamics in the bulk of these
alloys, especially in NiCoCr, NiFeCoCr and NiFeCoCrMn using spin polarized neutron scattering or muon spin rotation/relaxation.

\begin{acknowledgments}
The Compton scattering and soft x-ray absorption spectroscopy experiments were performed
with the approval of the Japan Synchrotron Radiation Research Institute (JASRI),
proposal numbers 2016B0131 and 2017B1243, respectively.
D.A.L. gratefully acknowledges the financial support of the National Secretariat of Higher Education,
Science, Technology and Innovation of Ecuador (SENESCYT).
S.M., G.D.S., G.M.S. acknowledge funding support by the Energy Dissipation and Defect Evolution (EDDE), an Energy Frontier Research Center funded by the U. S. Department of Energy (DOE), Office of Science, Basic Energy Sciences under contract number DE-AC05-00OR22725.
We gratefully acknowledge the financial support of the UK EPSRC (EP/R029962/1, EP/L015544/1 and EP/S016465/1).
\end{acknowledgments}


%

\begin{turnpage}
\begin{table*}[t!]
\caption{Magnetic moments from the orbital and spin sum rules (XMCD), KKR-CPA calculations (KKR), and bulk (SQUID and Compton) measurements.
The numbers in parentheses are statistical errors of one standard deviation at the magnitude of the least significant figure.
The statistical errors in the values derived from the bulk measurements are dominated by the
statistical error in the spin moment from Compton scattering so the errors in the total moments
from the SQUID have been omitted.
For NiFeCoCrPd, the average $m^{\rm orb}$ and average $m^{\rm spin,eff}$ from the sum rules were determined using their
respective Pd $d$-electron moments from the KKR-CPA calculations.
The Pd $d$-electron moments and those derived from them are indicated by asterisks.} 
\centering 
\begin{tabular}{c c c c c c c c c c c c c c } 
\hline\hline 
Alloy & Species & XMCD & XMCD & XMCD & XMCD & KKR & KKR & KKR & KKR & Bulk & Compton & Bulk & SQUID \\
& & $m_{z}^{\rm orb}$ & $m_{z}^{\rm spin,eff}$ & $m_{z}^{\rm orb}/m_{z}^{\rm spin,eff}$ & $m_{z}^{\rm tot,eff}$ & $m^{\rm orb}$ & $m^{\rm spin}$ & $m^{\rm orb}/m^{\rm spin}$ & $m^{\rm tot}$ & $m_{[100]}^{\rm orb}$ & $m_{[100]}^{\rm spin}$ & $m_{[100]}^{\rm orb}/m_{[100]}^{\rm spin}$ & $m_{[100]}^{\rm tot}$ \\
& & ($\mu_{\rm B}$) & ($\mu_{\rm B}$) & (no units) & ($\mu_{\rm B}$) & ($\mu_{\rm B}$) & ($\mu_{\rm B}$) & (no units) & ($\mu_{\rm B}$) & ($\mu_{\rm B}$) & ($\mu_{\rm B}$) & (no units) & ($\mu_{\rm B}$) \\
\hline 
NiFeCo     & Ni      &  0.087(9) &  1.1(1)  &  0.08(1)  &  1.2(1) &  0.0510 &  0.7008 &  0.0727 &  0.7518 &         &          &         & \\ 
           & Fe      &  0.053(5) &  2.5(3)  &  0.021(3) &  2.5(3) &  0.0601 &  2.5082 &  0.0239 &  2.5683 &         &          &         & \\ 
           & Co      &  0.16(2)  &  2.0(2)  &  0.08(1)  &  2.2(2) &  0.0854 &  1.6659 &  0.0512 &  1.7513 &         &          &         & \\ 
\hline 
NiFeCo     & Average &  0.10(1)  &  1.9(1)  &  0.054(5) &  2.0(1) &  0.0655 &  1.6250 &  0.0403 &  1.6905 & 0.02(2) & 1.64(2)  & 0.01(1) & 1.664 \\ 
\hline 
NiFeCoCr   & Ni      &  0.039(4) &  0.51(5) &  0.08(1)  &  0.55(5) &  0.0163 &  0.2736 &  0.0596 &  0.2899 &          &          &         & \\ 
           & Fe      &  0.049(5) &  1.8(2)  &  0.027(4) &  1.8(2)  &  0.0515 &  1.9146 &  0.0269 &  1.9661 &          &          &         & \\ 
           & Co      &  0.10(1)  &  1.1(1)  &  0.09(1)  &  1.2(1)  &  0.0563 &  1.0605 &  0.0531 &  1.1168 &          &          &         & \\ 
           & Cr      &  0.012(1) & -0.76(8) & -0.015(2) & -0.75(8) &  0.0066 & -0.6507 & -0.0101 & -0.6442 &          &          &         & \\
\hline 
NiFeCoCr   & Average &  0.050(3) &  0.66(6) &  0.075(3) &  0.71(6) &  0.0327 &  0.6495 &  0.0503 &  0.6822 & 0.017(4) & 0.231(4) & 0.07(2) & 0.248 \\
\hline 
NiFeCoCrPd & Ni      &  0.068(7) &  0.61(6) &  0.11(2)  &  0.67(6) &  0.0335 &  0.4227 &  0.0791 &  0.4561 &        &        &        & \\ 
           & Fe      &  0.035(4) &  2.3(2)  &  0.015(2) &  2.4(2)  &  0.0592 &  2.4295 &  0.0244 &  2.4887 &        &        &        & \\ 
           & Co      &  0.18(2)  &  1.6(2)  &  0.11(2)  &  1.8(2)  &  0.0825 &  1.4861 &  0.0555 &  1.5686 &        &        &        & \\ 
           & Cr      &  0.016(2) & -1.1(1)  & -0.015(2) & -1.1(1)  &  0.0069 & -1.0608 & -0.0065 & -1.0539 &        &        &        & \\
           & Pd      & 0.0069$^{*}$ & 0.1394$^{*}$ & 0.0491$^{*}$ & 0.1462$^{*}$ & 0.0075 & 0.1041 & 0.0720 & 0.1116 &  &   &         & \\
\hline 
NiFeCoCrPd & Average & 0.061(4)$^{*}$ & 0.73(6)$^{*}$ & 0.084(5)$^{*}$ & 0.79(6)$^{*}$ & 0.0379 & 0.6763 & 0.0560 & 0.7142 & 0.067(6) & 0.474(6) & 0.14(1) & 0.541 \\ 
\hline 
NiFeCoCrMn & Ni      &  0.029(3)  &  0.31(3)   &  0.09(1)  &  0.34(3)  &  0.0214 &  0.3059 &  0.0700 &  0.3273 &        &        &        & \\ 
           & Fe      &  0.029(3)  &  1.2(1)    &  0.023(3) &  1.3(1)   &  0.0621 &  2.1242 &  0.0292 &  2.1863 &        &        &        & \\ 
           & Co      &  0.090(9)  &  0.89(9)   &  0.10(1)  &  0.98(9)  &  0.0729 &  1.2046 &  0.0605 &  1.2775 &        &        &        & \\ 
           & Cr      &  0.0081(8) & -0.54(5)   & -0.015(2) & -0.53(5)  &  0.0087 & -1.2647 & -0.0069 & -1.2560 &        &        &        & \\
           & Mn      &  0.0084(8) &  0.0073(7) &  1.1(2)   &  0.016(1) &  0.0198 &  1.8415 &  0.0108 &  1.8613 &        &        &        & \\
\hline 
NiFeCoCrMn & Average &  0.033(2)  &  0.38(3)   &  0.09(4)  &  0.41(3)  &  0.0370 &  0.8423 &  0.0439 &  0.8793 &        &        &        & 0.008 \\
\hline \hline
\end{tabular}
\label{tab1}
\end{table*}  
\end{turnpage}

\end{document}